# Multi-shape memory by dynamic elastocapillary self-assembly

Dongwoo Shin and Sameh Tawfick

Mechanical Science and Engineering, University of Illinois Urbana-Champaign

Inspired by the synchronized beating of cilia,[1] we show that the collective dynamics of hair-like fibers in a meniscus during fast drainage enables their self-organization into multiple topologies including complex shape inversions. By draining liquid from triangular-base hair bundles, we demonstrate their transformations into concave hexagons, rounded triangles, circles and inverted triangles. These topologically distinct shapes are quenched collective mode shapes of the beating hair each corresponding to specific drainage rates of the liquid, and cyclic shape re-transformations can be simply stimulated by repeated immersion and drainage. The various topologies correspond to multiple elastocapillary equilibria. Complex cellular materials with varying pore size and density can be obtained by changing the drain rates from hair assemblies. Due to its simple implementation and energy efficiency, these shape transformations can have applications ranging from three-dimensional lithography to smart multi-functional surfaces.

Since their discovery, shape memory alloys have attracted our society for their use in artificial muscles, morphing airplane wings and medical stents.[2] They can be deformed and retain a specific temporary shape, but recover their original "memorized" shape by thermal or magnetic stimulation. The shapes are retained in both states even when the stimulus is removed. This fascinating behavior is enabled by the thermo-mechanical atomic re-organization between austenite and martensite equilibrium states. Likewise, polymers exhibit shape memory by activating molecular level bonding switches.[3] Effective shape recovery can be obtained by switching off and on polymer crosslinks at different states of mechanical strain. The shape recovery is entropically driven by the elasticity associated with the polymer chains' conformations and microstructural reorganization.[4] Moreover, by synthesizing network of molecular switches with more than one distinct transition temperatures, two-shape memories can be attained. A rich library of molecular switches currently exists to achieve thermal, photonic, magnetic and electric stimulations. Unfortunately, shape transformation has low efficiency of less than 5% due to heat losses and microstructural hysteresis, can achieve only small strains of less than 10%, and usually simple shape changes such as linear elongation or bending.

At small scale, capillary forces can bend or twist slender objects such as hair and has been used to engineer carbon nanotubes (CNTs) and nanopillars, a phenomenon called elastocapillarity.[5] Upon drying, surface forces fixate the shapes thus functioning as physical crosslink switches. Depending on the surface roughness, wettability, and fibers' stiffness, the hair's interfacial adhesion can persist, or debonding occurs as soon as the shape is dry.

We have observed new and interesting elastocapillary transformations of hair bundles as shown in **Figure 1**. Carbon fibers[6] of 5 μm diameter and 10-20 millimeter lengths[7] are organized normal to the substrate in a two-dimensional array in the form of an equilateral triangle of a few millimeters width. When the hair-like fibers are immersed in liquid (typically deionized water) followed by slow drainage or evaporation, the triangular bundle is transformed into a concave hexagon (CH) resembling a star, and upon drying, the shape is retained due to van der waals forces for months at ambient conditions. Upon re-immersion, the original triangular organization is immediately recovered (**Videos S1 - S3**). Further, we discovered that when the liquid is drained at higher rates, the hair bundle cross section can re-organize into a variety of shapes having distinct geometries ranging from rounded triangles (RT), circular (CL), three-lobed clubs (CB) and even inverted triangles (IT) as shown in **Figure 1**. Similarly, we tested hairs organized into a curtain of bundles having axes ratio 1:6 as a simple non-axisymmetric shape (**Video S4**). At slow drainage rate, the hairs self-organize into two lobes, and at high rates into an ellipse with switched axes. Thus, for triangles and ellipses, distinct shapes are obtained at specific drainage rates and remain fixed by surface forces. The observed shape transformations disrupt our understanding of elastocapillary self-assembly, and comprise all elements of a new self-programmable multi-shape memory effect stimulated by simple wetting and drying.

We found out that the shape transformations represent collective equilibria of the hairs stabilized by the competition between liquid capillarity and fiber bending. We constructed a mathematical model of 600 hairs arranged along the circumference of a triangular bundle. Each hair is considered as a slender rigid cylinder hinged at the substrate. The stiffness of the hinge is calculated from on the bending elasticity of the hair.[8] During drainage or drying, the wet hairs minimize their surface energy by aggregation, thus replacing the liquid-vapor interfaces surrounding each individual straight wet hair with the lower surface energy of bent coalescing hairs. To consider the collective mechanics of these hairs, we followed the static equilibria branches[9] representing each shape observed in the experiments as shown in **Figure. 2a**. We considered the mechanics of hair aggregation causing a decrease in area in the range of 0.8 to 0.3 depending on the initial fiber density. Because our bundles comprise a very large number of small fibers, the surface energy change between the wet and dry shape is $\sim \delta A . \gamma . \pi d . l$ where $\delta A$ is the normalized bundle area decrease, $\gamma$ the water surface energy, $\pi d$ the hair's perimeter and $l$ its length. The strain energy of the hairs $\sim 3EI\delta^2/l^3$ where $E$ is the hair's Young's modulus, $I$ is its second moment of area, and $\delta$ its displacement. A mode shape plot is constructed in **Figure 2b**, having the elastic strain energy per hair on the y-axis and the average change in surface energy on the x-axis.[10] The branch of the CH mode stems from the original triangular shape on the x-axis, showing that the equilibrium is achieved by reducing the surface energy while storing strain energy. The branch ends when static arrest[8a] is achieved, corresponding to fully-packed hairs. The slope of the plot represents the material properties such that for a given hair flexure stiffness and liquid surface tension, the equilibrium can be found from the intersection of the slope with the modes' branches. Another shape stemming from the triangle on the x-axis is the RT rounded triangle. This branch is shown for area changes in the same range of 0.8 to 0.3, and is constructed by gradually varying the radii of curvature at the corners (smallest curvature near the x-axis). At the end of this branch the equilibrium mode shape is the circular bundle having the diameter of the inscribed circle, which corresponds to areal shrinkage ratio of 0.6. Other RT branches can readily be constructed for shrinkage rate down to 0.3. Similarly, CL branch can be constructed for smaller final circles down to 0.3. Finally, the CB branch stems from the inscribed circle equilibrium and has three circular arcs. We also constructed the IT mode shape branch observed in the experiments by calculating the strain and surface energies associated with it. The plot conveniently predicts shape transformations of hair bundles under quasi-static drainage. The intersection of the slope with the first encountered branch represents the quasi-static equilibrium mode-shape, which agrees with our observations of CH obtained at very slow drainage rates and short lengths. The value of the slope increases with lower flexure stiffness (e.g. longer bundles) and higher liquid surface energy. As the liquid quasi-statically recedes, the hairs are quenched into the corresponding branch, and the mode shape is retained by surface adhesion when the liquid is fully evaporated. The intersection of the slope line with the following branch represents a higher mode shape.[11]

We systematically varied the length of triangular bundles and the drain speed to construct an experimental phase diagram for the shape transformations as shown in **Figure 3**.[12] Short bundles transform into CH at all speeds, which agrees with the mode-shape prediction. Snapshots of this transformation are shown in **Figure 3a**. At higher lengths, CH is obtained at low speed while CB is obtained at higher speeds. At lengths higher than 1.9 cm, the bundles transform from RT at low to IT at high drain rates. The higher mode shapes, in particular the IT, represent quenched topological states excited as a result of the finite fluid velocity. This dynamic hair re-organization can be understood in light of the two time scales corresponding to two characteristic frequencies: the meniscus capillary oscillation frequency $\omega_z$ and the flow-induced fiber oscillation frequency $\omega_x$. Lord Rayleigh showed that the period of capillary oscillations of droplets can be obtained for inviscid flow to be $T_z = (c_z \rho_L a_0^3/\gamma)^{1/2}$ where $c_z = 2\pi^2/3$ for the first mode of cylindrical meniscus or jet.[13] Using the inscribed circle (4.3 mm) as the size of the meniscus around the bundle, we calculate the theoretical time scale to be 86 ms, matching the period observed in the experiments for the shape transformation (**Video S5 and S6**). Notably, Bidone showed that when a free jet ejected from a triangular or elliptic orifice, an exchange between the kinetic energy of the liquid and the free surface tension leads to inverting the shape axes during capillary oscillations, which became known later as axis switching.[14] This is quite similar to the IT shape obtained in our experiments at high drain rate. In parallel, the drainage flow excites hair oscillations.[15] The hydrodynamic pressure on the fiber scales with $\rho_L v^2 \theta$ where $\rho_L$ is the liquid density, $v$ is the drainage velocity, and $\theta$ is the inclination angle of the fiber. This pressure induces the fiber motion and is hence balanced by the fiber's inertia $\rho_s d \omega_x^2 l \theta$ where $\rho_s$ is the solid density. The flow induced oscillations time scale is $T_x = 2\pi/\omega_x = (4\pi^2 \rho_s dl/\rho_L v^2)^{1/2}$.[16] By equating the two time scales, a critical velocity can be obtained as a function of length $v \sim (6\gamma \rho_s dl/\rho_L^2 a_0^3)^{1/2}$. We plot this relation on the experimental phase diagram of **Figure 3d** and it precisely predicts the drainage speed required to obtain high mode shapes.

These mode shapes manifest themselves as more complex topologies when hair is assembled into fractals of triangular bundles as shown in **Figure 4**. The fractals transform into cellular structures due to elastocapillarity. At low drainage rate, the fractal pores change their shape from triangular to polygons with the number of sides depending on the local boundary.[17] The mechanism of the transformation observed in the videos is shown in **Figure 4e** and is qualitatively similar to the quasi-static transformations of triangular bundles into CH. Surprisingly, at high drain rates, pores start disappearing until full-coalescence occurs into a single large CH. For example, a fractal of the six triangular bundles and three triangular pores can transform from a cellular structure with three self-organized pores, into two, one or no pores as the drain speed is increased (**Video S7**). Similarly, as shown in **Figure 4b**, a larger fractal with initially ten bundles and six pores, can transform between a cellular materials with six pores, three, two, one or no pores as the drainage speed is increased (**Video S8**). In both cases, these transformations are fully reversible, thus they represent stable shape memories easily accessible by varying the drainage speed. High-speed imaging confirms that the triangular pores become rounded yet remain filled with liquid at high drain rates (**Video S9 and S10**). The whole fractal undergoes first radial expansion, then, it contracts and the pores gradually shrink in size until they sequentially disappear.[18]

We speculate that the shape memory effect during the wetting and drying of hairs can be engineered for responsive surfaces with optical, texture, tactile, acoustic and mechanical functionalities.[19] It has clearly attractive advantages such as its simplicity, low cost and high energy efficiency compared to shape memory material. Additionally, the material system can be scaled down to nanometer scale as demonstrated in previous aggregation of CNTs.[20] While the higher dynamic modes were achieved in our work by controlling the drainage rate, there are many opportunities in decoupling the fiber oscillations from drainage or evaporation by using mechanical excitations. Perhaps this can become an example of borrowing a concept from nature, and doing much more with it.

**Acknowledgment**

This research is supported by the start-up funds from the Mechanical Science and Engineering at the University of Illinois Urbana-Champaign.

**Author Contributions**

S.T. designed the experiments, performed the mathematical modeling and wrote the manuscript. D.S. performed all the experiments, discussed the results and edited the manuscript.

**Competing financial interests**

The authors declare no competing financial interests.

**Multi-Shape Memory by Dynamic Elastocapillary Self-Assembly**

*Dongwoo Shin and Sameh Tawfick*
Mechanical Science and Engineering, University of Illinois Urbana-Champaign

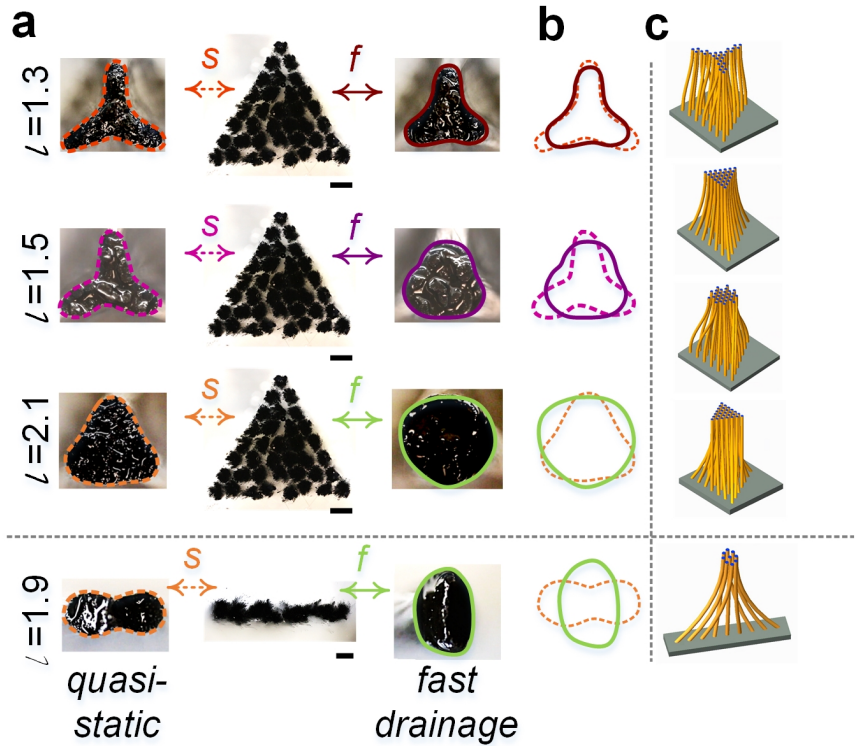

Figure 1. Dynamic elastocapillary shape transformations. (a) The left column shows the top view of hair bundle having various heights (*l*) after drying at slow drainage rate (s=0.01 cm/s), the middle column shows the bundles recovering their original arrangement when immersed back in the liquid, and the right column shows the bundle shapes at fast drainage rate (f=10 cm/s). Scale bar is 1 mm. (b) Schematics show the cross sectional changes for slow and fast drain rates. (c) 3D illustration of the hair show from top to bottom the concave hexagon (CH) arrangement, rounded triangle (RT), circle (CL), Inverted triangle (IT) and the axes switching of a linear curtain.

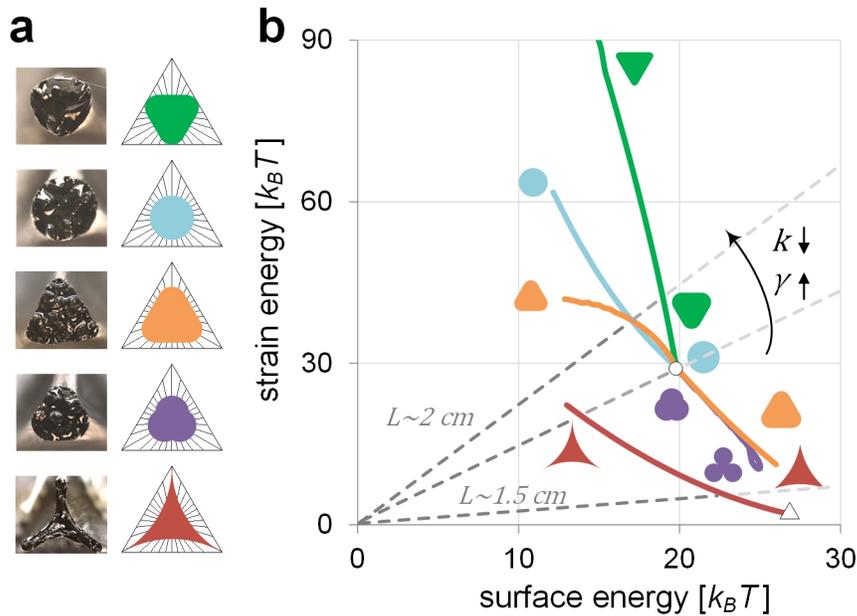

Figure 2. Elastocapillary equilibrium mode shapes of triangular bundles. (a) Optical images show the top view of bundle shapes (left) and their mathematical representation (right). Radial lines represent the displacement of individual fibers from the triangular base of the hair bundle to the new shape near the top. (b) Mode shape plot of elastocapillry equilibria of a triangular bundle numerically computed by considering the average surface and strain energy of 100 simulated fibers for each branch. The quasi-static shape can be obtained from the first intersection of the slope line with a branch. The slope increases as the bending stiffness decreases or the surface energy of the liquid increases. Higher modes are obtained at high drain rates.

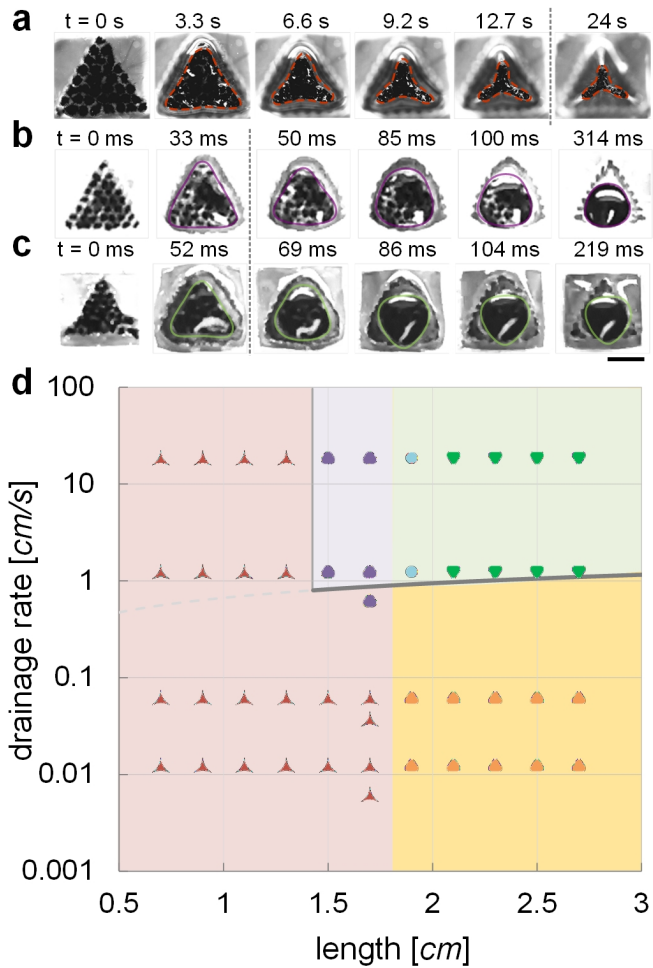

Figure 3. Dynamics time scale of elastocapillary shape memory. Optical images show snapshots of the top view of bundles during drainage for (a) 1.3 cm length bundle at 0.06 cm/s drain rate, (b) 1.5 cm length bundle at 18 cm/s drain rate, and (c) 2.1 cm bundle at 18 cm/s drain rate. Scale bar is 2.5 mm. (d) Experimental phase diagram showing the mode shapes obtained at various bundle lengths and drain rates. Symbol index is shown in Figure 2.

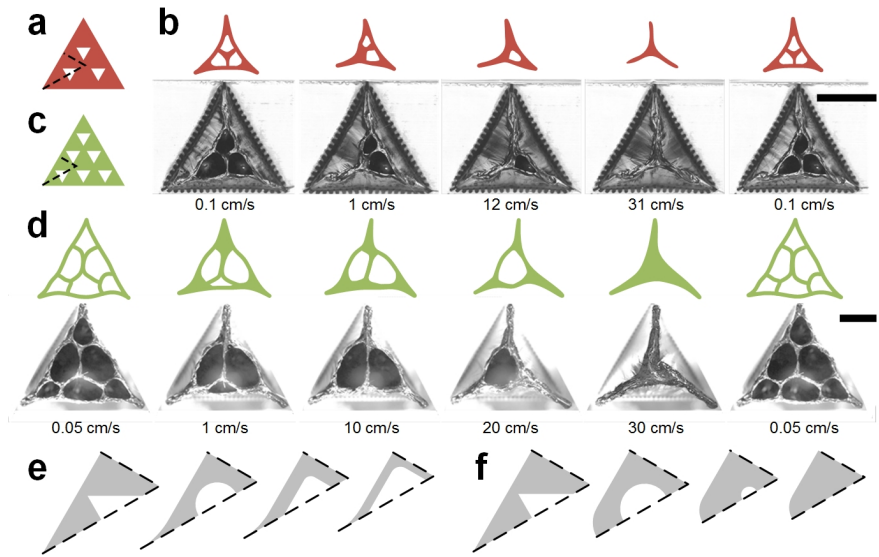

Figure 4. Complex multi-shape memory of triangular hair bundle fractals. Schematic of the design hair fractal bundles, where hair bundles are shown in red for small bundle design (a) and green for large bundle design (c). (b, d) Schematics and optical images show the top view of fractal bundles after drainage at various speeds: (b) four shape memories for small fractal design and (d) five shape memories for large fractal design. (e,f) Schematics of the fractal corners showing (e) the mechanism of pore shape transformation (white) at low drain rate and (f) the pore disappearance at high drain rate. Scale bars are 5 mm.